\documentclass[aps,pre,twocolumn,superscriptaddress,showpacs]{revtex4}
\usepackage{amssymb}
\usepackage{graphicx,color}
\usepackage{epstopdf}

\definecolor{r}{rgb}{1,0,0}   
\definecolor{g}{rgb}{0,1,0}   
\definecolor{b}{rgb}{0,0,1}
\makeatletter
\newcommand\footnoteref[1]{\protected@xdef\@thefnmark{\ref{#1}}\@footnotemark}
\makeatother

\begin{document}

%\preprint{}

\title{Penetration depth scaling for impact into wet granular packings}

% address or URL should go in the {}'s for \email and \homepage.

\author{T.A. Brzinski III}
%\thanks{}
\affiliation{
     Department of Physics \& Astronomy, University of
     Pennsylvania, Philadelphia, PA 19104, USA
}
\author{J. Schug}
%\thanks{}
\affiliation{
     Department of Physics \& Astronomy, University of
     Pennsylvania, Philadelphia, PA 19104, USA
}
\affiliation{
     William Penn Charter School,
     Philadelphia, PA 19144, USA
}
\author{K. Mao}
%\thanks{}
\affiliation{
     Department of Physics \& Astronomy, University of
     Pennsylvania, Philadelphia, PA 19104, USA
}
\affiliation{
     The Haverford School,
     Haverford, PA 19041, USA
}
\author{D.J. Durian}
%\thanks{}
\affiliation{
     Department of Physics \& Astronomy, University of
     Pennsylvania, Philadelphia, PA 19104, USA
}

%Collaboration name if desired (requires use of superscript address

\date{\today}

\begin{abstract}
We present experimental measurements of penetration depths for the impact of spheres into wetted granular media. We observe that the penetration depth in the liquid saturated case scales with projectile density, size, and drop height in a fashion consistent with the scaling observed in the dry case, but with smaller penetrations. Neither viscous drag nor density effects can explain the enhancement to the stopping force. The penetration depth exhibits a complicated dependence on liquid fraction, accompanied by a change in the drop-height dependence, that must be the consequence of accompanying changes in the conformation of the liquid phase in the interstices.
\end{abstract}

%45.70.-n Granular systems
%47.57.Gc Granular flow
%83.80.Fg Granular solids
%81.70.Bt Mechanical testing, impact tests, static and dynamic loads

%\keywords{}

\maketitle

%\section{}
%\subsection{}
%\subsubsection{}

% If in two-column mode, this environment will change to single-column
% format so that long equations can be displayed. Use
%\begin{widetext}
%\end{widetext}

\section{Introduction}

Many of the materials of interest in geophysics, mining, engineering, and industry are granular in nature. Most physics research has focused on dry, noncohesive grains, but in many situations grains are wet, either by circumstance or design. Introducing liquid to a granular packing results in dramatic, qualitative changes to its rheology, as can be observed when comparing the sand above and below the tide lines at the beach. Experimental studies of wet granular rheology have generally focused on static or low strain-rate bulk measurements~\cite{mitarai_wet_2006, strauch_wet_2012}. Measurements of angle of repose and angle of stability~\cite{Hornbaker_1997,Bocquet_2004,Nowak2005}, tensile strength~\cite{pierrat_1997,scheel_2008}, and yield under uniaxial compression~\cite{Iveson_2005}, all reveal a similar non-monotonic dependence on the fraction of pore space, $S$, occupied by the liquid phase: the stress at which a packing fails increases with $S$ for small $S$, and decreases as $S$ approaches unity, but exhibits little dependence for intermediate liquid fractions. The liquid fraction dependence of the repose angle in the limit of vanishing $S$ has been shown to be consistent with the balance between stresses due to the weight of the packing and the stress provided by liquid bridges between grains~\cite{Nowak2005}. Direct imaging of liquid conformation for small liquid fractions, in parallel with measurements of tensile strength, revealed that tensile strength saturates as the number of liquid bridges in the packing reaches its maximum~\cite{scheel_2008}. Thus statics experiments have shown that cohesive forces that arise from surface tension strengthen the packing. But as we consider dynamics there must also be viscous interactions between moving grains, and even the liquid inertia must become important for fast dynamics. Unfortunately such dynamical effects have been difficult to characterize with conventional rheometric techniques, which are frustrated by shear localization.

An alternative approach which has been successful for dry packings~\cite{RuizReview} is to study impact. Fig.~\ref{fig:apparatus} is a cartoon illustrating the geometry of a typical impact: a projectile falls from rest at height $h$ above a granular packing, and will come to rest after penetrating the packing to depth $d$. $H=h+d$ is the total drop distance, so by conservation of energy, the average force exerted by the packing on the projectile is $\langle F\rangle =mgH/d$, where $m$ is the projectile mass, and $g$ is gravity. Various empirical models have proven successful in describing the scaling of $d$ under certain conditions~\cite{uehara_low-speed_2003,walsh_2003,pica_ciamarra_dynamics_2004}. The case of shallow penetration for spherical projectiles of diameter $D$, free falling from rest onto a bed of hard, dry, non-cohesive grains was explored in \cite{uehara_low-speed_2003,ambroso_penetration_2005,nelson_projectile_2008}, and shown to have the form
% ******************EQUATION
\begin{equation}
	\label{eq:drypred}
	d=(0.14/\mu){\rho_{n}}^{1/2}D^{2/3}H^{1/3}~.
\end{equation}
% ******************END
Here $\rho_{n}=\rho_{p}/\rho_{g}$, where $\rho_{g} = \phi\rho_{gm}$ is the mass density of the granular media, $\phi$ is the volume fraction of grains in the packing, $\rho_{gm}$ is the mass density of the grain material, and $\rho_{p}$ is the density of the projectile; $\mu=\tan{\theta_R}$ is the internal coefficient of friction for the dry granular material, where $\theta_{R}$ is the angle of repose. For deeper penetration, $d$ scales as though the deceleration is constant~\cite{pica_ciamarra_dynamics_2004}, while for larger impact energies, $d$ has been observed to scale linearly with the momentum of the projectile~\cite{walsh_2003}. Volfson and Tsimring~\cite{tsimring_modeling_2005} showed that the $H^{1/3}$ scaling in Eq.~(\ref{eq:drypred}) is consistent with a modified Poncelet equation of motion, commonly used in high-speed ballistics~\cite{Allen1957,Backman1978,Forrestal1992}. In ref.~\cite{katsuragi_unified_2007} impact dynamics were reported for a steel sphere impacting a granular material with a wide range of speeds, 0-400 cm/s, over which $d\sim H^{1/3}$ holds.  The acceleration data were found to be consistent with a Poncelet-like equation of motion of the particular form:
%******************EQUATION
\begin{equation}
m a = -mg + F\left(z\right) + b v^2
\label{eq:ForceLaw}
\end{equation}
%******************END
where $m$ is the projectile mass, $g=9.8$~m/s$^2$, and \{$z, v, a$\} are respectively the projectile depth, velocity, and acceleration, and $F\left(z\right)$ is of the particular form $kz$.  The coefficient k and b depend on material parameters~\cite{brzinski_iii_characterization_2009,Katsuragi_2013} and projectile geometry~\cite{brzinski_now,Abe_2014}. Recent experiments have explored the dependence of $k$ and $b$ on the acoustic properties of the granular material~\cite{Abe_2012}, and identified a critical packing fraction, away from which the force law becomes non-linear due to changes in granular density during impact~\cite{Goldman_2010}. Eq.~(\ref{eq:ForceLaw}) with $F\left(z\right)=k z$ may be solved analytically for the penetration depth~\cite{brzinski_now, Katsuragi_2013}. While the result is consistent with with the $H$, $D$, $\rho$, and $\mu$ dependence of Eq.~(\ref{eq:drypred}), the mapping is not exact (see Fig.~5 of Ref.~\cite{Katsuragi_2013}). Over the full range of accessible parameters, where Eq.(1) is a good empirical description, both the inertial and rate-independent components of the stopping force must be taken into account:  upon initial impact $bv^2$ is at its largest while $k z$ vanishes, and vice-versa when the projectile comes to rest.

Though the models discussed above have yet to be rigorously tested and modified in order to describe wet materials, there have been several recent experimental studies of impact in wetted grains~\cite{Marston2012,KN2013,Takita2013}. Marston \textit{ et al.~}\cite{Marston2012} found that, for large impact velocities and small $S$, penetration depths can be greater than in the dry case. Only Marston \textit{et al.\ }have observed this increase in penetration depth: in all other cases, and for all liquid saturations, the penetration depth has been smaller than for a dry packing. Furthermore, in both Refs.~\cite{Marston2012}~and~\cite{Takita2013} there is no appreciable change in the $H$-dependence of the penetration depth on $S$. Nordstrom \textit{ et al.~}\cite{KN2013} study a system of grains fully submerged in a low-viscosity fluid. The authors again reproduce the $H^{1/3}$ scaling, and demonstrate that $b$ is proportional to the average mass density of the grain-fluid mixture.

Here we conduct a series of impact experiments in order to explicitly test the applicability of Eq.~(\ref{eq:drypred}) for shallow impacts onto granular packings as we vary liquid content from the dry case, $S = 0$, to the fully saturated $S = 1$ case. We recover a dependence of penetration depth on $S$ that is qualitatively similar to that in Ref.~\cite{Takita2013}, but also find that the presence of liquid in the packing changes the $H$-scaling of Eq.~(\ref{eq:drypred}). This is the first observation of this effect, which is non-monotonic in $S$. For both $S=0$~and~1 we recover $H^{1/3}$ scaling, but for intermediate values, we observe a higher-power relation. This power reaches a maximum between $S = 0.28$ and 0.43, and has a local minimum around $S = 0.56$. We also pay particular attention to the $S = 1$ case, where we vary drop height, projectile size, and material density, and find good agreement with Eq.~(\ref{eq:drypred}), only with a reduced coefficient. Because there are no liquid bridges between grains at $S=1$, one might expect hydrodynamic interactions to be the only cause for the reduced coefficient. We find that this cannot be the case, as the effect of the interstitial fluid is stronger for a less viscous fluid. We propose that dilatancy-induced perturbation of the liquid surface, and the resulting stress due to surface tension, account for the difference.

\section{Materials and Methods}

% ******************FIGURE
\begin{figure}%
	\includegraphics[width=0.75\columnwidth]{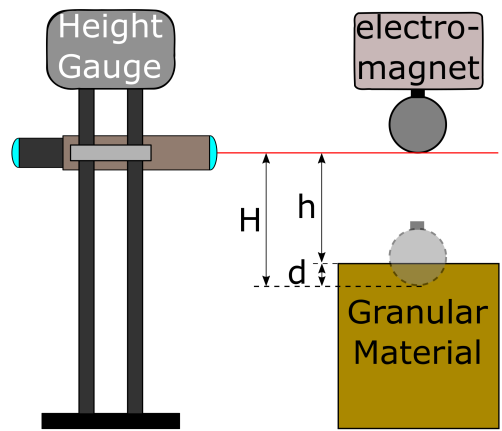}%
	\caption{A schematic of the experimental apparatus. The dashed, transparent image of the projectile demonstrates a typical example of penetration depth, which is always less than the projectile diameter in the present work. The red line indicates the point at which the free-fall height, $h$ is measured by the cathetometer before the impact. The penetration depth is $d$, and the total drop distance is $H=h+d$.}%
	\label{fig:apparatus}%
\end{figure}
% ******************END
% ******************TABLE
\begin{table*}[htbp!]
	\begin{minipage}{\textwidth}
		\caption{Material properties for the different bulk granular materials used, as well as the projectile diameters, projectile materials, projectile densities and liquid saturations tested for each material. Experiments conducted with water are by J.S. and T.A.B., all others are by K.M. and T.A.B.}
		\begin{tabular}{lccccccccc}
			\\ \hline\hline
			Fluid & $\eta_{f}$~[cP] & $\rho_{f}$~[g/cm$^3$] & $\sigma$~[dyn/cm] & $d_{g}$~[$\mu$m] & $S$ & $D$~[cm] & $\rho_p$~[g/cm$^3$] & projectile material & $\theta_{c}$ \\ \hline
			Air & 0.019 & 0.0013 & N/A & 365 & 0 & 2.54 & 0.464 & wood & N/A\\
			Oil & 29 & 0.85 & $32.5\pm2.5$ & 365 & 0~-~1 & 2.54 & 0.464 & wood & $26^{\circ} \pm 2^{\circ}$ \\
			Water & 0.89 & 1 & 72 & 515 & 1 & 1.27~-~5.08 & 0.685 & wood & $20^{\circ} \pm 2^{\circ}$ \\
			Water & 0.89 & 1 & 72 & 515 & 1 & 2.54 & 0.685~-~7.762 & miscellaneous & $20^{\circ} \pm 2^{\circ}$ \\ \hline\hline
			\label{tab:Materials}
		\end{tabular}
	\end{minipage}
\end{table*}
% ******************END

In the present work, all data are for impacts into packings of spherical glass grains wetted either by water or mineral oil (McMaster-Carr Supply Company, Light Viscosity). For all water-wetted samples, an open-topped, semi-transparent plastic beaker, 10.8~cm in diameter, is filled with 515~$\mu$m diameter glass spheres (Potters Industries, stock number P-0230). The water is poured by hand along the wall of the container and  into the granular packing. The pour rate is adjusted in situ such that no water accumulates at the packing surface, and so that the interface between wet and dry granular material is horizontal except near the pouring location. The water eventually fills to above the height of the granular material, and is allowed to overflow from the container. Finally, the container is tapped and squeezed by hand to deform the packing, thus releasing trapped air. This squeezing continues until no bubbling is observed at the packing surface, and no bubbles are visible against the semi-transparent wall of the container except near the bottom of the sample. As a final check, the wetted packing is weighed to ensure the mass is consistent with a fully saturated packing. While it is possible that some small air bubbles remain trapped in the bulk, the relatively large grainsize and slow pour-rate are intended to mitigate the formation of such bubbles. Note that all samples wetted with water are fully saturated, such that $S=1$. For all oil-wetted samples, we use 365~$\mu$m diameter glass spheres (Potters Industries, stock number P-0170), cleaned according to the recipe in Ref.~\cite{yuli}. Both the grains and the oil are massed, then stirred together by hand with a wooden spatula until the material has no clumps, and is homogenous to the eye. The resulting oil-wetted granular mixture is used to fill the plastic cylinder to above the lip, then leveled by scraping excess material off the top such that the surface of the granular material is flush with the lip of the container. The mineral oil has a viscosity of 32 times that of water, 0.45 the surface tension of water, and is 0.85 the density of water. The contact angle, $\theta_{c}$, of the oil and water on the granular material was determined to be $20^{\circ}$ for water and $26^{\circ}$ for oil. The contact angle for water was determined using the procedure described in Ref.~\cite{yuli}, whereas the value for oil was obtained by imaging an oil droplet on clean glass.  The liquid saturation, $S$, is determined by the relative masses of granular material and liquid, which we vary as indicated in Table~\ref{tab:Materials}.

The experimental apparatus is depicted in Fig.~\ref{fig:apparatus}. The projectiles are spherical, and both projectile radius and density are varied as reported in Table~\ref{tab:Materials}. Density is varied by using spheres of different material: wood, as well as various plastics and metals. Varying the projectile material also changes the surface roughness and wettability of the projectile, but we believe the mechanics of the system should be independent of these quantities. Stopping force should be independent of projectile roughness since it is dominated by frictional interactions between grains in the bulk (as opposed to between grains and the projectile)~\cite{brzinski_now}. Likewise, the stopping force should be independent of the wetting properties of the projectile because the stopping force is exerted primarily through interaction with the granular material, not the liquid interface. All projectiles have a small square rod protruding perpendicular to the projectile surface. The rod is capped with a piece of ferrous metal. The rod is a lightweight plastic, and the cap is of a size such that the mass of this assembly is less than the uncertainty in the mass of the sphere. The sphere is suspended by the ferrous cap from an electromagnet centered above the cylindrical container. We measure the height above the granular surface, $h$, from which the projectile is to be dropped with a height gauge and cathetometer (Titan Tool Supply, model TC-II). The projectile is released by turning off the electromagnet, allowing the projectile to fall and impinge upon the granular surface at impact speed $\sqrt{2gh}$. We measure the total distance through which the projectile falls, $H$, again with the height gauge. The penetration depth is then given by the difference $d=H-h$. In the present work, impacts result in shallow penetration, such that $d<D$.

A typical penetration depth is illustrated in Fig.~\ref{fig:apparatus}. None of impacts onto wetted grains discussed in this work resulted in a penetration much greater than the radius of the projectile, and even for impacts onto dry grains, all penetration depths are less than $D$. This is a consequence of limits to our dynamic range in $h$, but, conveniently, is also consistent with impacts for which, in dry systems, system size effects may be neglected. Because $d$ is always approximately an order of magnitude smaller than the diameter of the packing, our impacts are \emph{well} within the shallow regime for which the Janssen effect may be neglected. Furthermore, it's been demonstrated in dry packings that an impactor begins to interact with the sidewalls only for projectile-wall gaps less than a projectile diameter~\cite{nelson_projectile_2008}. In the case of our container, this limit corresponds to a projectile of diameter $D=3.6$~cm dropped onto the center of the packing. We are within that limit for 5 of the 7 projectile sizes discussed here. Additionally, the small penetration depths mean that even for those largest two projectiles, only two impacts of the largest ($D=5.08$~cm)projectile produced craters greater than 3.6~cm in diameter: one of diameter 4.52~cm, and the other of diameter 3.64~cm. Thus, we believe that any finite size effects in our system are negligible, except perhaps for the deepest single impact of the largest projectile.

\section{Results}
\subsection{Fully Saturated}

% ******************FIGURE
\begin{figure*}[htp]
	\begin{minipage}{\textwidth}
		\includegraphics[width=\textwidth]{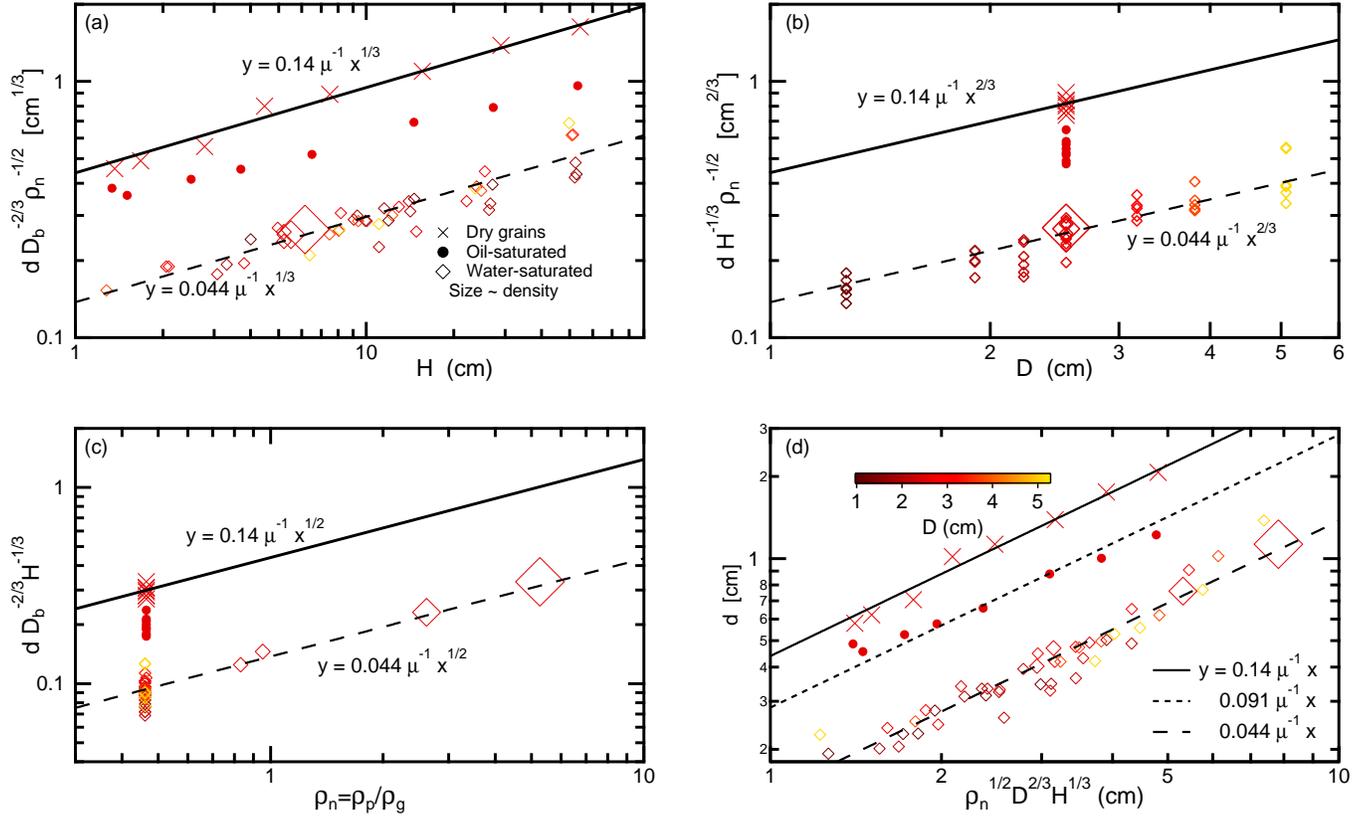}
		\caption{(Color online) Data for spherical projectiles of diameter 1.27, 1.91, 2.22, 2.54, 3.18, 3.81, and 5.08~cm (indicated by color), and density 0.685, 1.23, 1.41, 3.86, and 7.76~g/mL (indicated by symbol size), dropped onto packings of dry (crosses) and \emph{fully}-saturated ($S=1$) grains using each preparation (circles and diamonds for oil-saturated and water-saturated respectively). The solid lines are Eq.~(\ref{eq:drypred}). a)~$d D^{-2/3}{\rho_{n}}^{-1/2}$ plotted against total drop distance, $H$. The dashed line is a power law with a power of $1/3$. b)~$d H^{-1/3} {\rho_{n}}^{-1/2}$ plotted against projectile diameter, $D$. The dashed line is a power law with a power of $2/3$. c)~$d H^{-1/3} D^{-2/3}$ plotted against $\rho_{n}$. The dashed line is a power law with a power of $1/2$. d)~$d$ plotted against ${\rho_{n}}^{1/2}D^{2/3}H^{1/3}\mu^{-1}$. The dashed line is a fit of the water-saturated data to a proportionality. }
		\label{fig:stacked}
	\end{minipage}
\end{figure*}
% ******************END

We begin by examining the case of fully saturated packings, $S=1$, in which the interstitial volume is occupied exclusively by either water or mineral oil. In this case there are no interstitial air/liquid interfaces, and thus no cohesive interparticle forces due to surface tension. For comparison, data are also included for impact into dry grains. Measurements of penetration depth, $d$, for impacts of multiple projectiles dropped from a range of heights are plotted in Fig.~\ref{fig:stacked} in order to compare with Eq.~(\ref{eq:drypred}). The data are plotted against each of our control parameters in turn: (a)~the total drop distance, $H$; (b)~projectile diameter, $D$; (c)~normalized projectile density, $\rho_{n}$. In Fig.~\ref{fig:stacked}(a), the penetration depth is normalized according to the $\rho_{n}$ and $D$ scaling predicted by Eq.~(\ref{eq:drypred}), yielding $d D^{-2/3}{\rho_{n}}^{-1/2}$. Surprisingly, the penetration depth is not only collapsed by this normalization, but also scales, as in the dry case, as $H^{1/3}$ for each liquid-saturated packing. Following the same prescription for Fig.~\ref{fig:stacked}(b), the penetration depth is normalized by the expected $H$ and ${\rho_{n}}$ dependence to yield $d H^{-1/3}{\rho_{n}}^{-1/2}$. While projectile size and density were varied only in the water saturated case, we find again that the normalization collapses the data, and that the projectile size dependence is consistent with the $D^{2/3}$-scaling of Eq.~(\ref{eq:drypred}). Likewise, in Fig.~\ref{fig:stacked}(c) the data are collapsed by the normalization $d H^{-1/3} D^{-2/3}$, and demonstrate ${\rho_{n}}^{1/2}$ scaling, consistent with Eq.~(\ref{eq:drypred}).

To coalesce Figs.~\ref{fig:stacked}(a-c), we plot penetration depth, $d$ as a function of ${\rho_{n}}^{1/2}D^{2/3}H^{1/3}$ in Fig.~\ref{fig:stacked}(d). We observe that all three data sets collapse onto lines of slope~$1$, reflecting that the penetration depth scaling for impacts into liquid saturated grains is consistent with Eq.~(\ref{eq:drypred}). However the constant of proportionality is different for each packing preparation: $0.14$~for dry grains, $0.091$~for oil-saturated grains, and $0.044$~for water-saturated grains. Thus, Eq.~(\ref{eq:drypred}) captures the full dependence of $d$ on total drop distance $H$, projectile size $D$, and material density $\rho_{n}$ for both dry and saturated packings, but the empirical prefactor is dependent on some property or properties of the liquid: viscosity, surface tension, or fluid density. 

The scaling behavior of the data in Fig.~\ref{fig:stacked}(d) for the liquid-saturated packings is the same as for dry packings, except for a numerical pre-factor, which suggests that Eqs.(\ref{eq:drypred},\ref{eq:ForceLaw}) are relevant.  But since they include neither viscosity nor surface tension terms, these forces seem unlikely to play a role. The participation of viscous forces is further counterindicated by the fact that penetration is greater for oil than for water, while the viscosity of oil is more than an order of magnitude greater than that of water. Also, the grain size used for oil saturated grains is smaller, which should result in larger relative viscous forces.

Unlike viscosity, the role of fluid density can be argued from Eq.~(\ref{eq:drypred}) and the materials dependence of the coefficients in Eq.~(\ref{eq:ForceLaw}). There are two effects. First, $bv^2$ represents inertial drag.  So $b$ should be proportional to the total mass density $\phi\rho_{gm}+(1-\phi)\rho_f$ of the grain plus fluid mixture.  A nonzero $\rho_f$ gives a larger inertial drag and hence a smaller penetration. While Eq.~(\ref{eq:ForceLaw}) cannot be mapped onto Eq.~(\ref{eq:drypred}), we suppose that a liberal estimate for this effect is to modify the granular medium density in Eq.~(\ref{eq:drypred}) from $\rho_g=\phi\rho_{gm}$ to $\rho_g=\phi\rho_{gm}[1+(1/\phi-1)\rho_f/\rho_{gm}]$.  Second, $kz$ represents quasistatic friction due to gravitationally loaded grain-grain contacts~\cite{brzinski_now}. Interstitial fluid unloads these contacts due to buoyancy, and this should cause a smaller quasistatic stopping force and hence a larger penetration. Thus we expect $k\propto \mu (\rho_{gm}-\rho_f)$, and suppose this may be accounted for by modifying the effective friction coefficient in Eq.~(\ref{eq:drypred}) to $\mu(1-\rho_f/\rho_{gm})$.  The repose angles of dry and submerged grains differ by only a couple degrees, so $\mu$ itself is nearly unaltered~\cite{wet_repose}.  Altogether then, the effect of nonzero fluid density would be to change the numerical coefficient of Eq.~(\ref{eq:drypred}) from 0.14 to $0.14/[(1-\rho_f/\rho_{gm})\sqrt{1+(1/\phi-1)\rho_f/\rho_{gm}}]$.  This yields 0.19 and 0.21 for oil- and water-saturated packings, respectively. However this is not consistent with our observation of shallower penetrations, with measured coefficients of 0.091 and 0.044. By this estimate, the reduced friction, not the increase in inertial drag, dominates the mechanics, but the estimated density effect is opposite to that observed.  While the density effect thus cannot explain our results, it could play a role in the deeper penetrations observed in Ref.~\cite{Marston2012}. A more likely scenario is that the so-called ``lubrication effect" observed in Ref.~\cite{Marston2012} is the consequence of the granular phase collapsing from a low volume-fraction to a more compacted state. Indeed, Marston \textit{ et al.\ }observe that the effect becomes weaker if, during the preparation of the granular material, the container is tapped. This explanation fits well with the packing fraction dependence observed by Umbanhowar and Goldman~\cite{Goldman_2010} for dry packings.

Finally, the static packing is not cohesive in the case of $S=1$, so it seems unlikely that surface tension could be responsible for the reduced penetration depths demonstrated in Fig.~\ref{fig:stacked}. However, the surface tension of water is significantly higher than that of mineral oil, consistent with deeper penetration into oil-saturated packings than into water-saturated packings. It may be that impact stimulates surface tension-driven cohesion in the bed. One possible mechanism is suction as a result of a reduction of pore pressure during impact due to dilatancy and capillary action at the packing boundary. The smaller contact angle for grains in water could lead to a relative enhancement of such transient capillary effects in water-saturated beds. This mechanism would most likely modify the friction-like term of the stopping force by enhancing contact forces between grains, but if this is the case, it is surprising that $d$ demonstrates scaling consistent with Eq.~(\ref{eq:drypred}). Capillarity with the projectile could also arise. The only way to entirely eliminate such capillarity effects would be to conduct impact experiments with an entirely submerged system.

\subsection{Partially Saturated}

% ******************FIGURE
\begin{figure}
\includegraphics[width=\columnwidth]{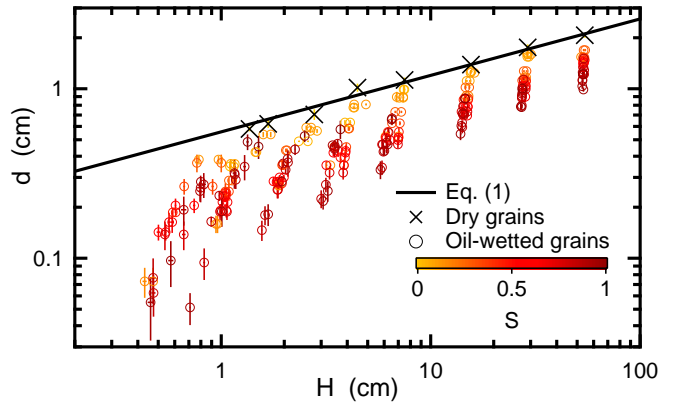}
\caption{(Color online)  Penetration depth, $d$, plotted against total drop distance, $H$, for spheres of diameter $2.54$~cm and density 0.464~g/mL, dropped onto a bed of spherical glass beads, both dry (black crosses), and wetted by oil (circles), with liquid saturations as indicated by color.}
\label{fig:raw}
\end{figure}
% ******************END

Next we observe the effects of varying the amount of interstitial liquid by conducting impacts onto packings which are partially saturated with oil. Penetration depth, $d$, as a function of total drop distance, $H$, is plotted in Fig.~\ref{fig:raw} for spheres dropped onto granular packings of different liquid saturations, $S$. Spheres dropped onto packings of dry grains penetrate as Eq.~(\ref{eq:drypred}), in agreement with previous experiments. \emph{All} spheres dropped onto packings of wet grains stop at a depth shallower than for spheres falling through the same total distance into dry packings, indicating that a wet packing will always exert a stopping force greater than in the dry case. Additionally, the data deviates from the $H^{1/3}$~scaling observed for dry packings.

\begin{figure}
\includegraphics[width=\columnwidth]{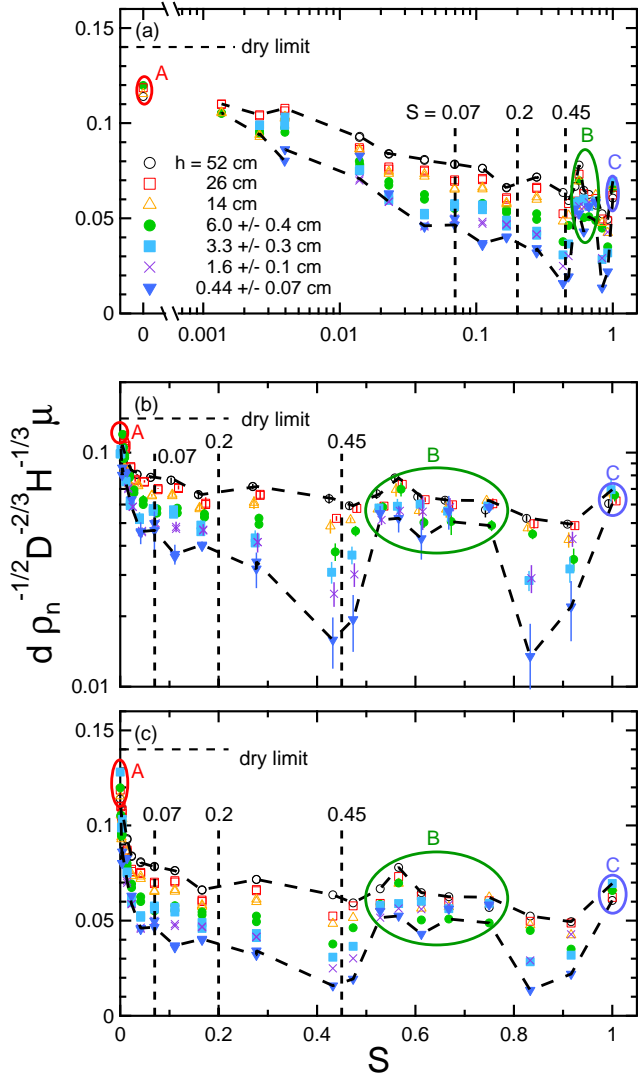}
\caption{(Color online) Penetration depth, $d$, normalized according to Eq.~(\ref{eq:drypred}), plotted against liquid saturation for spheres of diameter $2.54$~cm and density $0.68$~g/mL, dropped into a bed of spherical glass beads wetted by oil, for various drop heights, $h$, as indicated by color. The data points for the largest and smallest normalized penetration depths at each value of $S$ are connected by solid black lines to highlight variations in the quality of collapse. Data are plotted on (a)lin-log, (b)log-lin and (c)linear axes in order to highlight features at different scales. Labels~A-C and the vertical dashed lines indicate features of the data and values of $S$, respectively, which are discussed in the text.}
\label{fig:satlog}
\end{figure}

To understand how these two effects vary with liquid saturation, Fig.~\ref{fig:satlog} shows $d$, as a function of $S$ for several drop heights, where $d$ is normalized according to Eq.~(\ref{eq:drypred}). We see the same general behavior for all values of $H$: the reduced penetration depths are non-monotonic with $S$. There is a rapid initial reduction of penetration depth between $S=0$ and $S=0.07$, followed by a range of $S$-values for which $d$ decreases slowly. At approximately $S = 0.45$, $d$ reaches a local minima, then increases to an approximate, normalized value between 0.05 and 0.08, varying little till $S=0.8$. Finally, we see $d$ decrease to another local minima, then increase once more as the packing approaches submersion at $S=1$.

% ******************FIGURE
% ******************END

Note that in Fig.~\ref{fig:satlog}, as in Fig.~\ref{fig:stacked}, normalizing according to Eq.~(\ref{eq:drypred}) collapses the data for both $S=0$ (pt.~A) and $S=1$ (pt.~C). However, for intermediate values of $S$ the data collapse fails, and the $H$-scaling of penetration depth demonstrates a strong, non-linear $S$-dependence. The relation between $d$ and $H$ becomes dramatically stronger than a $1/3$~power law as $S$ increases from $0$ to approximately $0.07$, coincident with the rapid initial decrease of penetration depth. The spread of the penetration data fluctuates a bit between $S=0.05$ and $0.45$, but remains approximately constant. For $0.5<S<0.8$ (label~B), the effect becomes very small, with $d$ once again approaching $H^{1/3}$ scaling, though the data collapse is still worse than at pts.~A and C. The data collapse fails once again for $0.8<S<1$, before collapsing once more as $S\to 1$.

While there is a temptation to relate these behaviors to the distribution of liquid in the pores, the lack of data on liquid conformation in 3-D renders such statements purely speculative. That said, some excellent data on liquid distribution for $S\le0.3$ was presented in Ref.~\cite{scheel_2008}, where it is observed that, as $S$ increases, the number of liquid bridges increase as well. Beyond $S=0.07$, liquid bridges start to coalesce, with the coalesced liquid network exhibiting a percolation transition around $S=0.2$. Both of these values are indicated by dashed vertical lines in Fig.~\ref{fig:satlog}, and the first point, $S=0.07$, coincides well with both the end of the rapid reduction of $d$, and with the saturation of stronger $H$-dependence. This suggests that cohesion due to surface tension at air-liquid interfaces is likely to be the primary mechanism by which the liquid contributes to the stopping force for small values of $S$. Neither the percolation transition at $S=0.2$, nor the subsequent evolution of the liquid conformation seems to have a dramatic effect on $d$ till approximately $S=0.45$. Presumably for $0.45<S<1$, the dominant fluid mechanism transitions from capillary action across liquid bridges to whatever mechanism enhances the stopping force at $S = 1$, but this transition is very non-monotonic. Of particular interest is the near-collapse of the data between the two local minima in penetration depth, labeled pts.~B in Fig.~\ref{fig:satlog}. For this range of liquid saturations, there is still air-liquid interface within the pore space, yet the $H$-dependence of $d$ approaches the $1/3$ power law observed for dry and saturated systems.

\section{Conclusions}

In conclusion, we find that the penetration depth, $d$, of a projectile impacting upon a granular packing is always reduced when an interstitial liquid is added to the packing. For fully liquid-saturated packings the functional dependence of $d$ on $\rho_{n}$, $D$, and $H$ is the same as Eq.~(\ref{eq:drypred}) for dry grains. The unchanged $H$- and $D$-dependence of the penetration depth reflect that the rate- and intruder size-dependence of the stopping force is the same as for the dry case, suggesting that the enhanced stopping force is not due to a viscous contribution. This interpretation is borne out by the fact that we measure deeper impacts for the more viscous fluid (oil). This result indicates that the enhanced stopping force is the consequence of a more complex mechanism, perhaps involving surface tension acting at the boundaries of the packing. The wettability of our granular materials is consistent with this interpretation: our water-wetted grains have a lower contact angle, which ought to enhance surface tension effects in that system.

In contrast to the saturated case, for $0<S<1$, Eq.~(\ref{eq:drypred}) no longer describes the functional dependence of $d$ on $H$. In fact, the $S$-dependence of penetration depth is strongly non-monotonic. Thus, while the depth-averaged stopping force exerted by a granular packing upon the projectile is enhanced by the presence of an interstitial fluid under all conditions, which is the dominant mechanism behind this enhancement must be dependent on $S$. The dramatic increase in stopping force for small amounts of liquid, $S<0.07$ and subsequent $S$-independence tracks well with reports of the total area of liquid-air interfaces in the bulk~\cite{scheel_2008}, indicating that this initial enhancement of the stopping force may be due largely to capillary action across liquid bridges. We are not aware of any theoretical or structural data to compare to for $0.45<S<1$, but as $S\to 1$, capillary action between individual grains must vanish. Both the intervening non-linearities in penetration depth and the changes in functional form indicate that there is a complex evolution of the liquid conformation before the pores reach saturation.

In the future, either experimental data or theory describing the conformation of liquid in the pores could illuminate the complexity we observe in the impact dynamics for $0.45<S<1$. For all values of $S$, direct measurements of the stopping force and high-resolution position data taken \emph{during} impact might enable us to determine the full form of the stopping force, mirroring the approach we have used to study the case of $S = 0$~\cite{uehara_low-speed_2003, tsimring_modeling_2005, katsuragi_unified_2007, brzinski_now}. Also, while it is difficult to vary viscosity and surface tension independently, experiments with a broader variety of liquids would help to differentiate the roles of viscosity, density and surface tension. To test the idea that dilatancy plays an important role at $S=1$, high speed imaging of the packing surface during an impact may be sufficient. If not, tools like X-ray imaging could be used to directly measure the liquid fraction in situ. In the present work, we assume boundary effects or of a scale similar to that observed in dry experiments, but that is not necessarily the case: the effect of interstitial liquid on ball-wall interactions is certainly of experimental interest. Lastly, our results for fully-saturated grains indicate that wettability of the granular material may dramatically alter the granular response to impact. We would expect such an effect to be particularly strong for partially saturated packings. The most obvious way in which wettability might be important is by modulating the force that can be exerted on the grain-scale due to surface tension. Furthermore, the spatial distribution of the liquid within the pores must depend on wettability, so the ensemble effect may be even greater. Thus, an impact study in which wettability is systematically varied, as in Ref.~\cite{yuli}, would be of great interest.

\begin{acknowledgments}
We would like to thank Jeremy Marston and Kerstin Nordstrom for helpful discussion. This work is supported by the National Science Foundation through grants MRSEC/DMR-{1120901} and DMR-{1305199}.
\end{acknowledgments}

%\bibliographystyle{aps}
\bibliography{Manuscript}

\end{document}